# OPTIMIZATION OF DD-110 NEUTRON GENERATOR OUTPUT FOR BORON NEUTRON CAPTURE THERAPY

HOSSAM DONYA[1,2], Muhammed Umer[1]

[1]Physics Department, King Abdulaziz University, Jeddah 21589, Saudi Arabia

[2]Physics Department, Faculty of Science, Menoufia University, Shebin El-Koom, Egypt

## ABSTRACT

Boron neutron capture therapy is about a century old, but still current and active. In this treatment, a high absorption cross-section of boron is used for thermal neutrons and the excited state is decomposed into alpha and lithium ions. These daughter particles have a high LET value and transfer their energy to the surrounding cells. In this work, the neutron generator DD-110 is the neutron source of the treatment mode. Since the neutron generator produces fast neutrons, these must be thermalized to the energy range of the epithermal neutrons. The beam shaping assembly is used for this purpose. The effectiveness of the system is analyzed using in-air and in-phantom parameters. The system provides an acceptable ratio of thermal to epithermal flux, gamma dose to epithermal flux and beam collimation. However, the epithermal flux and the ratio of fast neutron dose to epithermal flux need to be further improved. This can be achieved by improving the performance of the neutron generator to 2E13 n/sec. The entire system is modeled and simulated with the MCNPX code.

## INTRODUCTION

One kind of radiation treatment that makes use of the special qualities of boron-10, a stable isotope of boron, is boron neutron capture therapy (BNCT). Certain cancers appear to respond well to BNCT, especially those that are in the brain and other delicate regions where conventional radiation therapy may be difficult to administer.[1-3] The idea behind BNCT is that atoms of boron-10 absorb thermal neutrons and subsequently break down producing particles with a high LET. High-energy lithium and alpha ions are produced by this process, and the near-cancer cells primarily absorb them.[4] This differential effect results from the fact that alpha and lithium-ion have short route track lengths and that boron-10 preferentially accumulates in tumor cells at higher concentrations than in normal cells. Then within a cell's dimensions, they run out of energy.

$^{10}B + n \text{ (thermal)} \rightarrow {}^{11}B \rightarrow {}^{4}He \text{ (1.78 MeV)} + {}^{7}Li \text{ (1.01 MeV)}$ (6.3%)

$^{10}B + n \text{ (thermal)} \rightarrow {}^{11}B \rightarrow {}^{4}He \text{ (1.47 MeV)} + {}^{7}Li \text{ (0.84 MeV)} + \gamma \text{ (0.48 MeV)}$ (93.7%)[5]

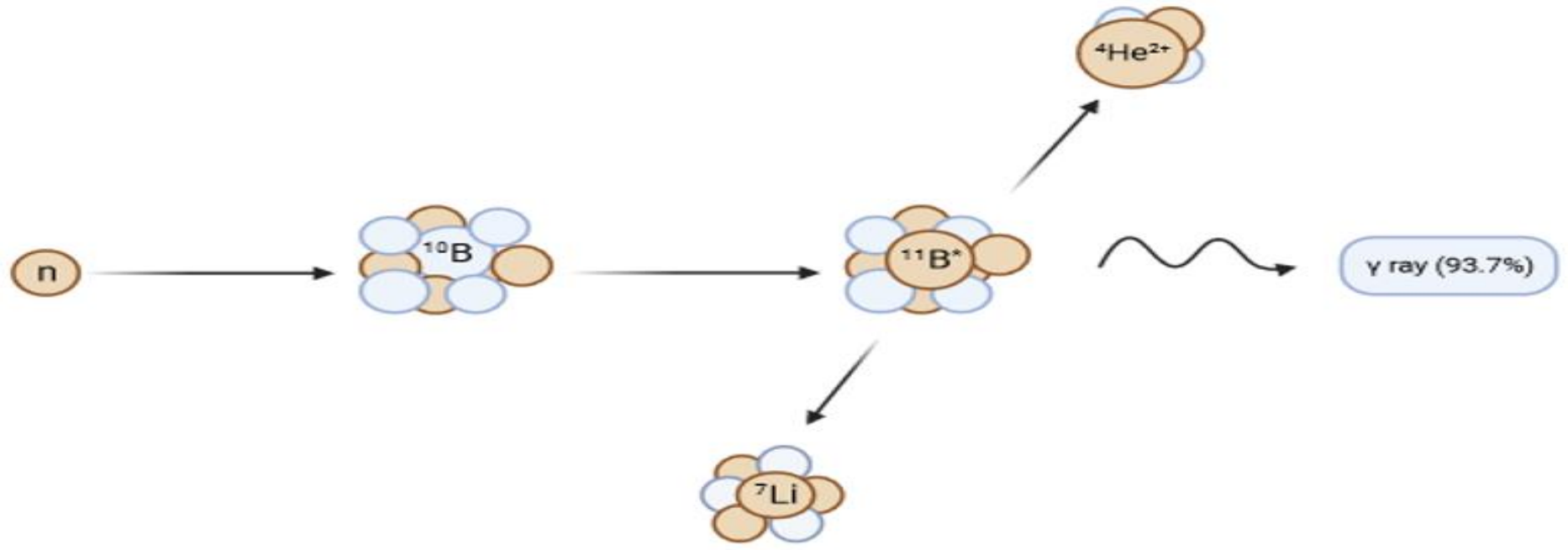

*Figure 1 Boron Neutron capture nuclear reaction (93.7 %)*

An injectable substance containing boron-10 is first administered to the patient in BNCT administration. This chemical accumulates in the tumor due to high metabolism rate of tumor cells compared to normal tissue cells. Once the patient has amassed enough of the substance within the tumor, high-energy particles that cause damage to the tumor cells are released when the patient is exposed to a neutron beam.

Compared to traditional radiation therapy, which uses an x-ray or electron beam, BNCT has the advantage of employing high LET particles, attaining a larger absorbed dosage to the tumor relative to normal tissue, and preserving normal tissues.[6]

It was 1936 when Gordon Locher first proposed BNCT. He put forward the principles of BNCT following the insight of Goldhaber. Goldhaber discovered that the stable $^{10}B$ element has a large thermal neutron capture cross section and then immediately breaks down into an energetic $^{4}He^{2+}$ with a rebounding $^{7}Li$ ion in 1934. These particles have a total average kinetic energy of 2.33 MeV and a range of 12–13 µm in tissue, which is comparable to the 10–100 µm cellular dimensions of mammals. During 1959 - 1961, the first BNCT clinical examinations were carried out at Brookhaven National Laboratory (BNL) and Massachusetts General Hospital (MGH) to treat Glioblastoma Multiforme (GBM).[7] Borax, or sodium tetraborate, was the boron delivery agent. Like most scientific discoveries, the initial outcomes were not satisfactory. Following years of a lot of advancements, there are currently 33 BNCT facilities globally, spread across 13 countries.[7]

The neutron source, neutron energy spectrum, and boron delivery agent have the greatest influence on the effectiveness of BNCT. Because of its strong affinity for tumor cells, 4-borono-L-phenylalanine (BPA) is now the most employed boron agent in BNCT.[8] Since the efficiency of the treatment is dependent on the ratio between tumor boron concentration to normal tissue boron concentration and this is ever improved, this area is still active to enhance this ratio and to decrease the toxicity of the boron compound. Reactors and particle accelerators are the options that provide the right neutron flux to reduce treatment time when it comes to neutron sources for BNCT. Reactors are the original BNCT neutron sources and provide higher neutron fluxes than particle accelerators[9]; however, Reactor-based BNCT (RB-BNCT) is very expensive to install, difficult to operate, and requires a license because it is not registered as a medical device.[1] Therefore, the current neutron sources for BNCT are particle accelerators. For neutron sources, there are various kinds of particle accelerators and reactions. These sources include deuterium accelerators for the reactions $^{7}Li(^{2}H,n)^{8}Be$, $^{9}Be(^{2}H,n)^{10}B$ and $^{13}C(^{2}H,n)^{14}N$, as well as proton accelerators with targets of Li - $^{7}Li(p,n)^{7}Be$ or Be - $^{9}Be(p,n)^{9}B$. The other accelerator is a deuterium accelerator for the fusion reaction of d + d and d + t. In this work, Deuterium accelerators for fusion reaction, also known as neutron generators, are selected as the source of neutron.[10]

The nuclear processes resulting in 14.1 Mev and 2.45 Mev, respectively, are d + t - 3H(d,n)4He and d + d - 2H(d,n)3He in the neutron generators. The tritium element is a beta emitter, and the d + t process produces far more energetic neutrons than epithermal neutrons, hence it is unlikely to be selected as a BNCT neutron source.[10] In this work d + d neutron generator is utilized as a neutron source and especially DD-110 neutron generator, which is built from Adelphi tech., is employed. There is no previous work that assesses this neutron generator model as a neutron source for deep-seated BNCT. Beam shaping assembly (BSA) is intended to attenuate the 2.45 Mev neutrons, that this generator produces, down to the epithermal region (0.5 eV to 10 keV). The reason the BSA's output port detects epithermal neutrons instead of thermal ones is that, as they approach deeply ingrained tumors, their energy levels drop, and they transform into thermal neutrons which are then captured by the boron element. Using the MCNPX code, the optimization of the output neutrons through BSA is investigated. The in-air and clinical parameters of the overall design are also evaluated using MCNPX.

## METHODS AND MATERIALS

### DD-110 Neutron Generator

The DD-110 neutron generator's design is the initial step in this project. An ion beam from a microwave plasma ion source drives the generator. The ion source generates a high plasma density for a high current and high D+ content by utilizing the electron cyclotron resonance (ECR) effect. It accelerates to the deuterium target after the deuterium ion is created. The target is 50 mm in diameter. Neutrons are isotopically expelled from the collision when the deuterium target is bombarded by the accelerating deuterium. Monoenergetic, 2.45 Mev, $1 \times 10^{10}$ neutrons per second are released in this system. The whole size of the neutron generator is 26.67 cm diameter and 63.5 cm length cylinder.[11]

### Beam shaping assembly

The design of the beam shaping assembly that thermalizes the fast neutrons to epithermal neutrons is the next stage. Table 1 lists the classification of neutrons according to their energy.

| Neutron type | Energy range |
|---|---|
| Thermal neutron | < 0.5 ev |
| Epithermal neutron | 0.5 ev – 10 kev |
| Fast neutron | > 10 kev |

*Table 1 energy-based neutron classification [10]*

The goal of BSA is not only to reduce neutron energy; it also aims to direct the isotopic neutrons towards the BSA output port, reduce the energy of fast neutrons and in turn their number; filter out gamma rays produced by the collision of fast neutrons with hydrogen and nitrogen atoms; shield the surrounding environment and reduce out-of-field exposure; and collimate the final yield to the appropriate aperture dimension. Table 2 provides a numeric guideline for these elements.

| Parameter | Recommended value |
| --- | --- |
| Therapeutic epithermal flux ($\varphi_{epth}$) | $\geq 5 \times 10^{8}$ $n_{epith}.cm^{-2}s^{-1}$ |
| Thermal to epithermal flux ratio ($\varphi_{ther}/\varphi_{epth}$) | $\leq 0.05$ |
| Beam collimation (J/$\varphi_{epth}$) | $\geq 0.7$ |
| Fast neutron dose to epithermal flux ratio ($D_{fast}/\varphi_{epth}$) | $\leq 7 \times 10^{-13}$ Gy.cm$^2$ |
| Gamma dose to epithermal flux ratio ($D_{\gamma}/\varphi_{epth}$) | $\leq 2 \times 10^{-13}$ Gy.cm$^2$ |

*Table 2 IAEA recommendation of neutron beam parameters for BNCT [10]*

The IAEA has proposed these values as the beam spectrum standards for epithermal neutron-based treatments. The BSA is built with parts that can provide these outcomes to meet these requirements. The moderator, which reduces the fast neutron energy to an epithermal neutron energy range, the reflector, which returns the neutrons that are outgoing from the collimator direction, the filter, which removes unwanted beam components, and the collimator, which collimates the final output to the desired region, are the key parts of the BSA. The general setup of the neutron generator and BSA looks like in Figure 2.

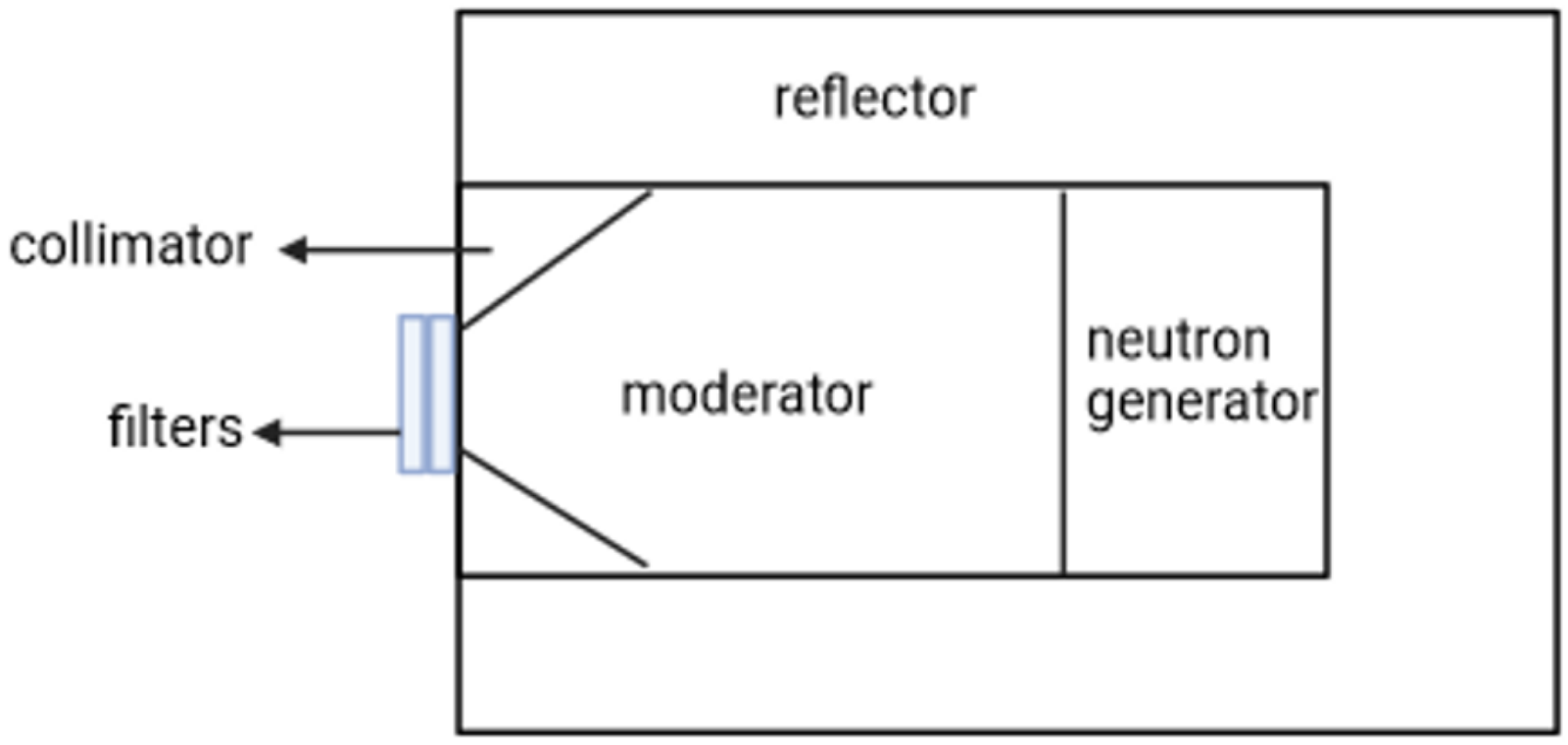

*Figure 2 General beam shaping assembly (BSA)*

The first step of designing BSA is to select the appropriate moderator material and its dimensions. Moderator material is selected based on the ability of elastic collision with neutrons. So they are mostly lightweight materials. Further consideration must be given that the moderator does not decrease the neutron flux so much. The other parameter of the moderator is the thickness of the material. In this work, 10 materials as a moderator are assessed for the given neutron generator by using the MCNPX code.

Finding a suitable filter and reflector comes next after selecting the right moderator. The unique property of reflectors is being heavy material. The last step in designing the BSA is to research several collimator materials with various aperture sizes and select the best one. The choice of reflector and shielding for the neutron generator are made simultaneously. Subsequently, the aggregate impact of the several constituents of BSA is evaluated.

## Clinical parameters and Dose calculation

After the overall design of the BSA, its clinical efficacy should be studied by using phantoms. Mostly used in-phantom parameters for BNCT are Advantage depth (AD), Advantage ratio (AR), Advantage depth dose rate (ADDR), and Treatment time. Advantage depth is a depth in the irradiated body at which the total therapeutic tumor dose equals the maximum healthy tissue dose. It is a measure of beam penetration ability. The advantage ratio is the ratio of the total therapeutic dose to the total healthy tissue dose over the AD. The advantage depth dose rate, as the name indicates, is the dose rate at the AD. Treatment time is the maximum irradiation time until the healthy tissue dose limit is reached.

Neutrons interact with atoms in 3 main ways. The first one is the elastic scattering of fast and epithermal neutrons by low atomic mass atoms. In the BNCT modality this type of reaction takes place when the neutron interacts with a hydrogen atom $^{1}H(n, n')^{1}H$, the second interaction is the inelastic scattering of neutrons. The last one is the neutron absorption reaction. This type of reaction is the main reaction that takes place in BNCT. Neutron absorption reactions in BNCT are $^{1}H(n, \Upsilon)^{2}H$, $^{14}N(n, p)^{14}C$, and $^{10}B(n, \alpha)^{7}Li$. Hence the absorbed dose in this treatment modality are these four types of reactions. These dose transfer mechanisms have different biological effectiveness. The overall calculation of the total absorbed dose is given in Equation 1.[12]

$$D_t = CBE_B D_B + RBE_H D_H + RBE_N D_N + RBE_\gamma D_\gamma$$

$CBE_B$ is compound biological effectiveness of boron dose, $RBE_H$, $RBE_N$, and $RBE_{\gamma}$ are relative biological effectiveness of hydrogen, nitrogen, and gamma doses respectively. Their values are given in Table 3.

| $CBE_B$(normal tissue) | $CBE_B$(tumor) | $RBE_H$ | $RBE_N$ | $RBE_{\gamma}$ |
|---|---|---|---|---|
| 1.35 | 3.8 | 3 | 3 | 1 |

*Table 3 radiation weight factors in BNCT [10]*

The boron dose analysis is dependent on the concentration of the boron compound in the tumor and the healthy tissue. In this work, the boron concentrations in the tumor and the normal tissue are 33.3 ppm and 10 ppm. The ratio between tumor boron concentration and healthy tissue boron concentration is 3.33.[13]

## RESULTS AND DISCUSSION

### Design of BSA

For the moderator, there are nine material candidates in this work: $MgF_2$, LiF, $D_2O$, $TiF_3$, Al, $AlF_3$, $Al_2O_3$, Pb, and $CaF_2$. Initially, these materials were examined using a standard reflector, devoid of a filter and collimator. Figure 3 shows their affinity to convert fast neutrons into epithermal neutrons at varying thicknesses.

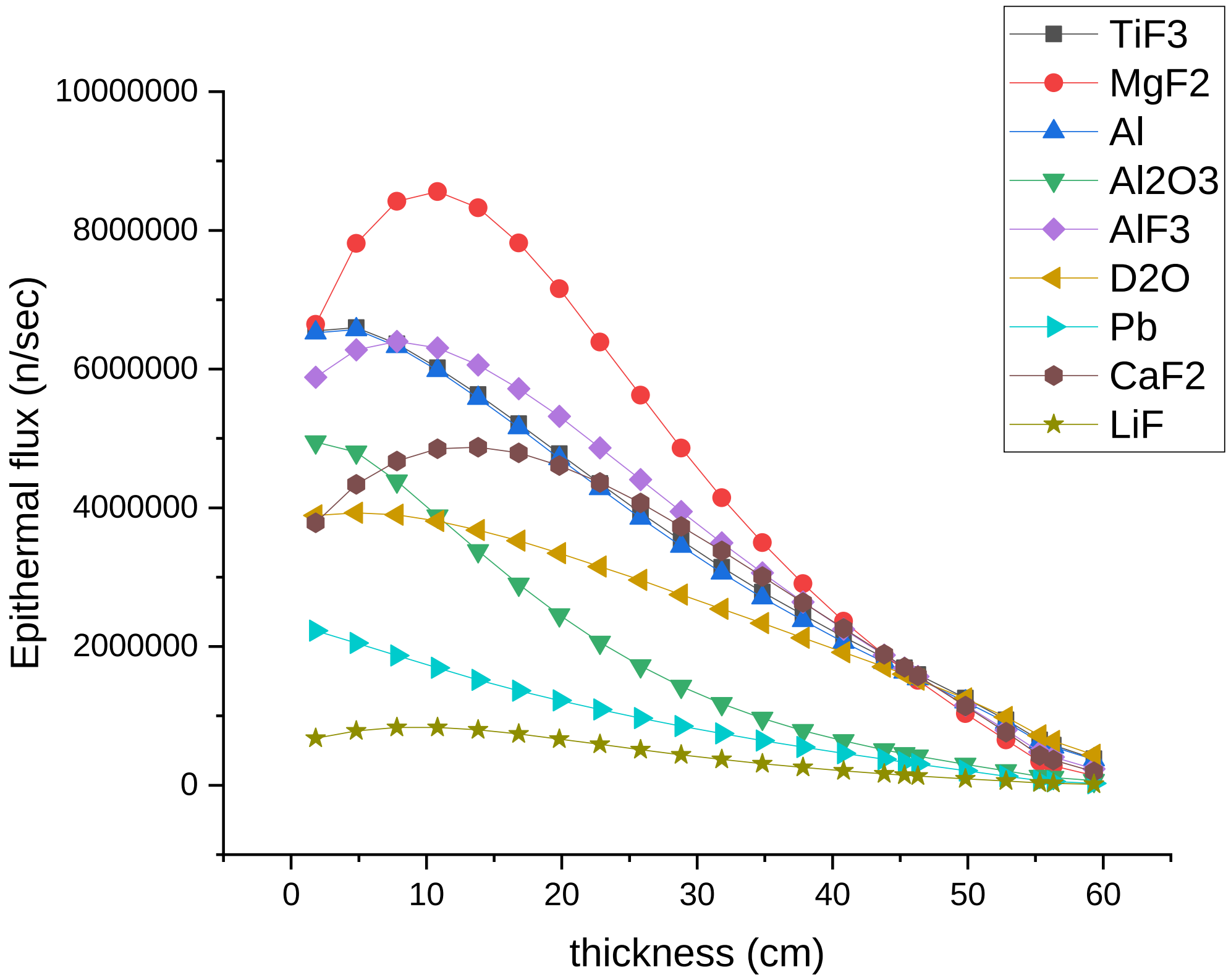

*Figure 3 Epithermal flux for different material moderators versus their thicknesses*

With a step of 3 cm, the epithermal flux is counted from 1.81 cm to 59.31 cm. The 2.45 Mev neutrons are isotopically expelled from the target, causing some of them to travel in different directions. The neutrons are then reflected by the reflector. They expend some energy on their travel. This explains why, for some materials, the epithermal flux at 1.81 cm is in the order of 1E06, which was an unexpected value. This graph makes it evident that the maximal epithermal flux and associated thickness of various materials vary. Certain materials exhibit more epithermal flux for a given neutron spectrum at a certain thickness in comparison to one another, but not across the entire thickness range. To demonstrate this, MgF2 exhibits the maximum epithermal neutron flux between 1.81 cm and 42 cm, but not at thicknesses greater than 42 cm. The reason for this is that, in addition to cross-section values for materials being energy dependent, the energy of the neutron changes as it passes through a material. As a result, the epithermal flux relative values for various materials differ. In most of the thicknesses, $MgF_2$, $TiF_3$, Al, $AlF_3$, and $CaF_2$ achieve better than the others, as advised by the IAEA report.[10] The epithermal flux is nearly constant for LiF and Pb. This is a positive result since they effectively filter out thermal, fast, and gamma rays without compromising the epithermal flux. Furthermore, materials with a high slope of degradation like $Al_2O_3$ are challenging to work with for optimizing other parameters.

The next step after choosing the best moderator is to reduce the doses of gamma, thermal, and fast neutrons using various moderator thicknesses and filters. In this project, filter candidate materials are LiF, Pb, and Bi; LiF is good at absorbing thermal neutrons, Pb is good at filtering gamma rays and fast neutrons, and Bi is a gamma ray filter. The effect of these materials in the BSA is displayed in Figures 4–6. The output epithermal flux decreases as the thickness of the filter material increases.

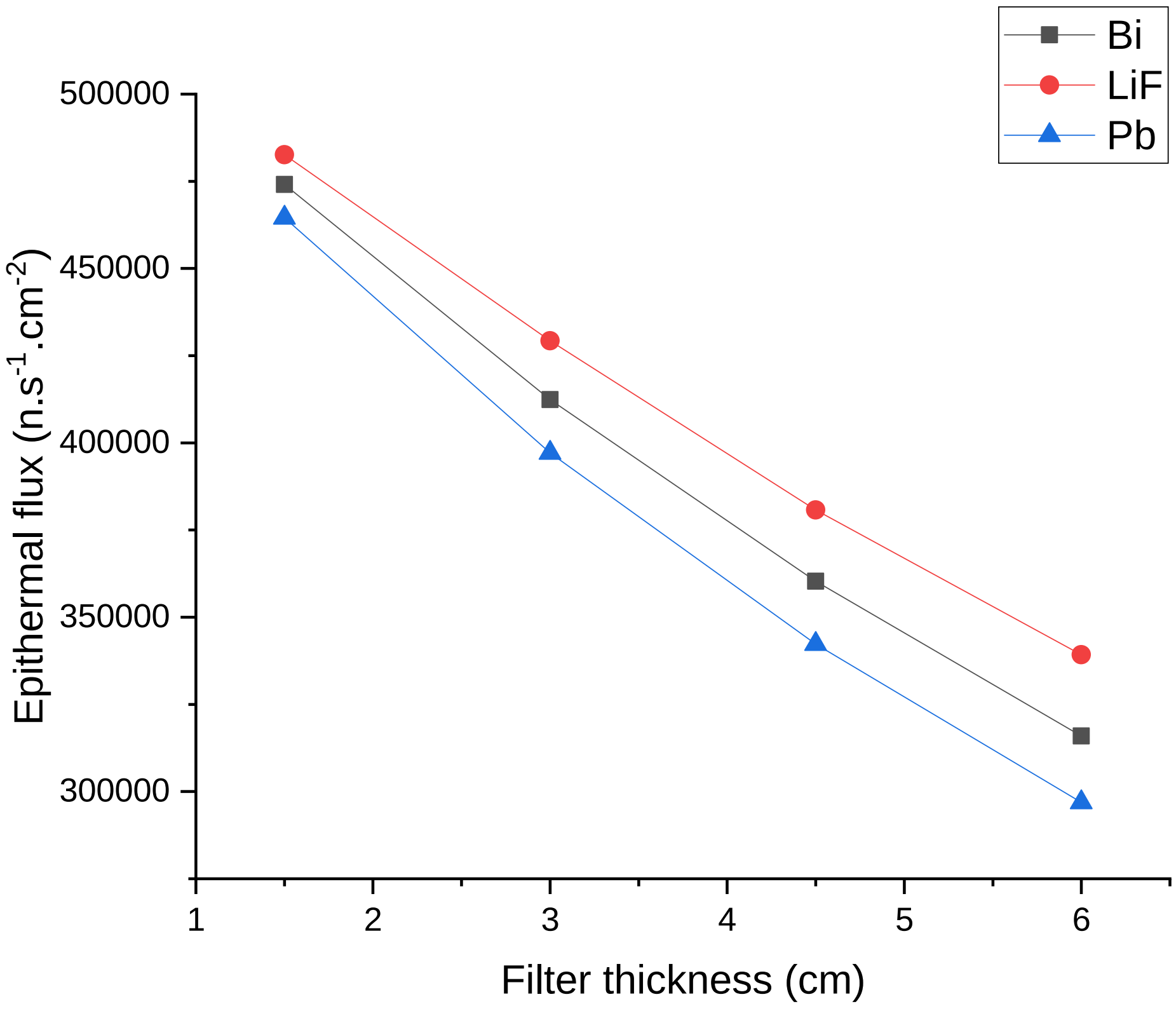

*Figure 4 Effect of different filter materials on epithermal flux of the BSA with moderator of 32 cm MgF2.*

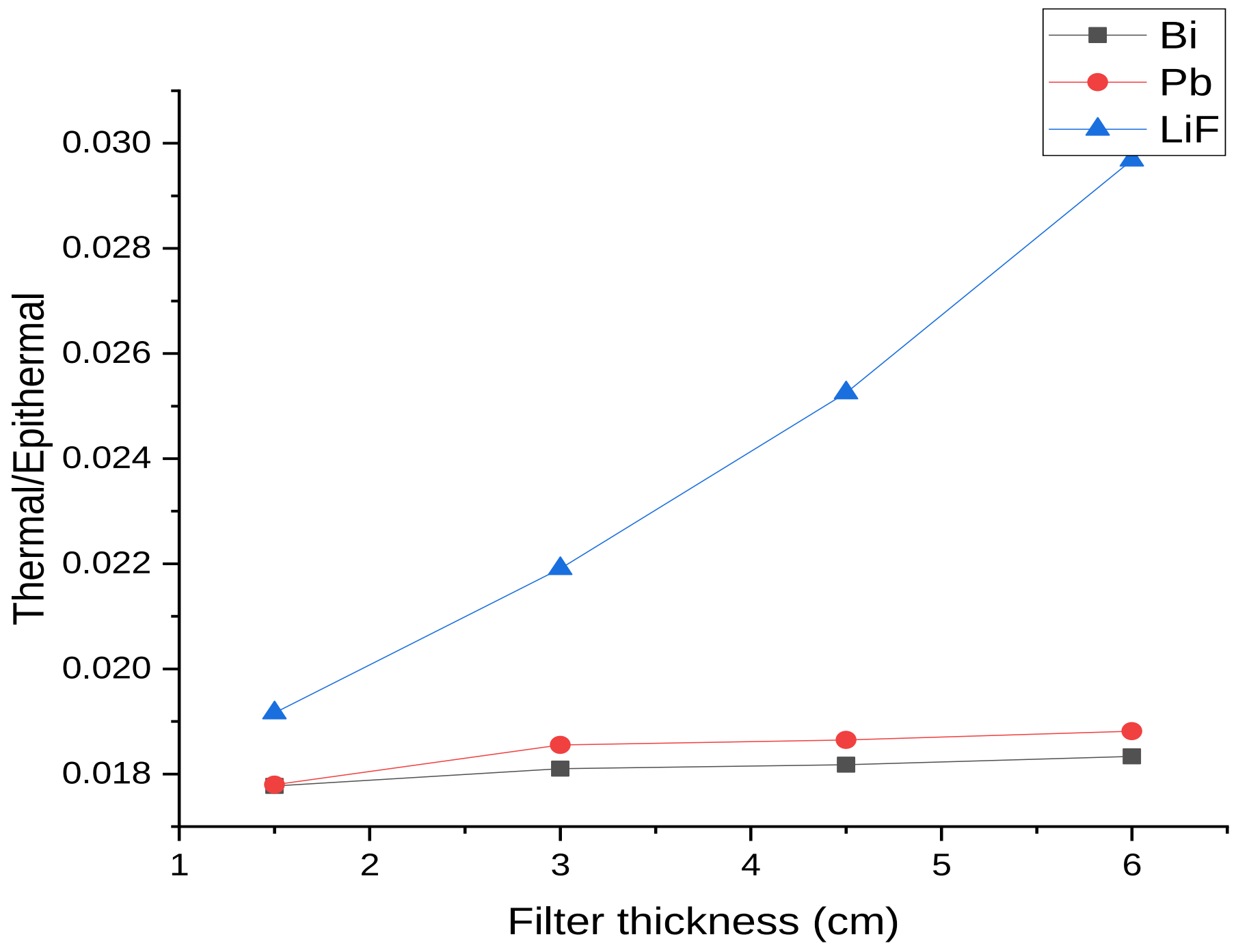

*Figure 5 Effect of different filter materials on thermal to epithermal flux ratio of the BSA with moderator of 32 cm MgF2*

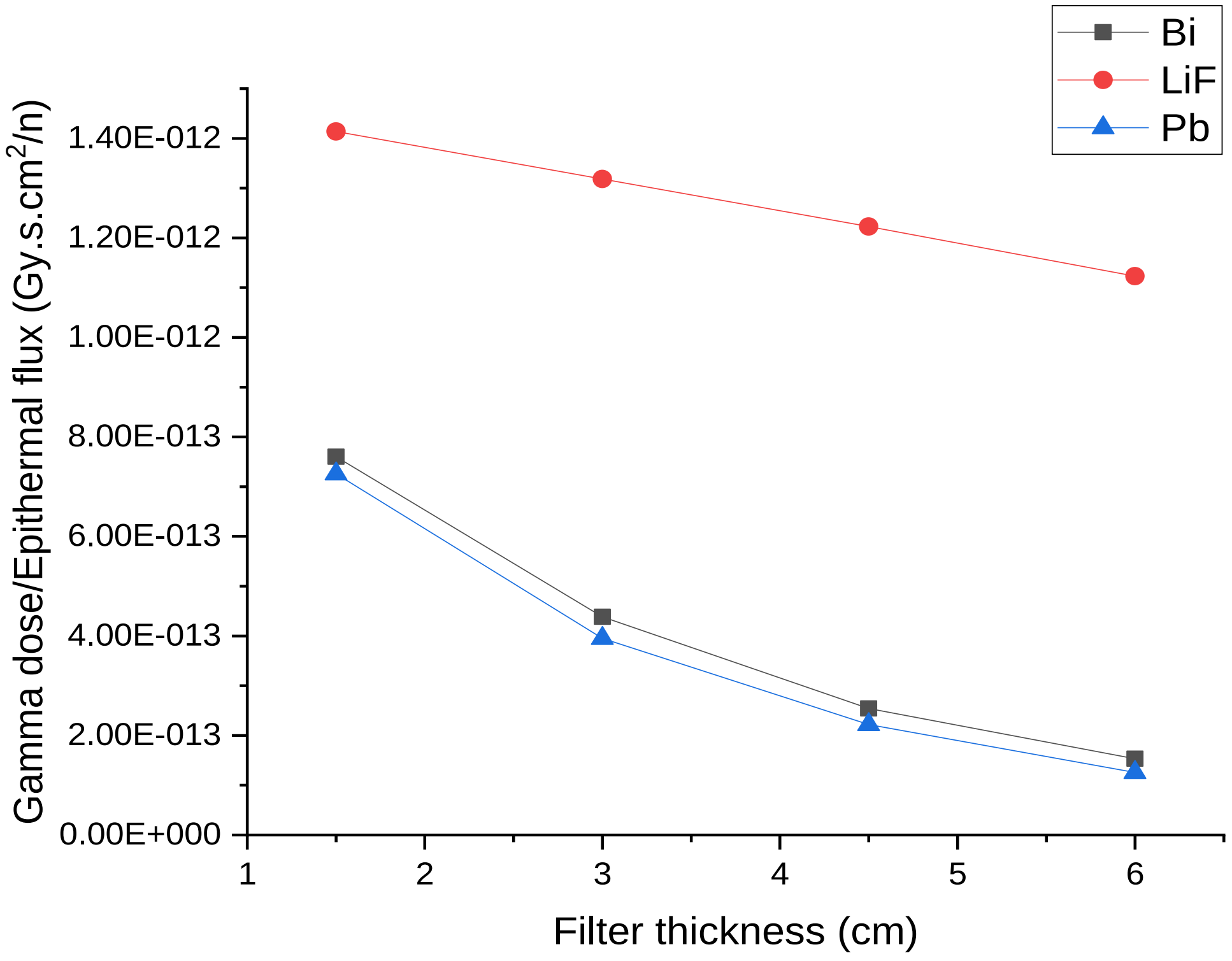

*Figure 6 Effect of different filter materials on gamma dose to epithermal flux ratio of the BSA with moderator of 32 cm MgF2*

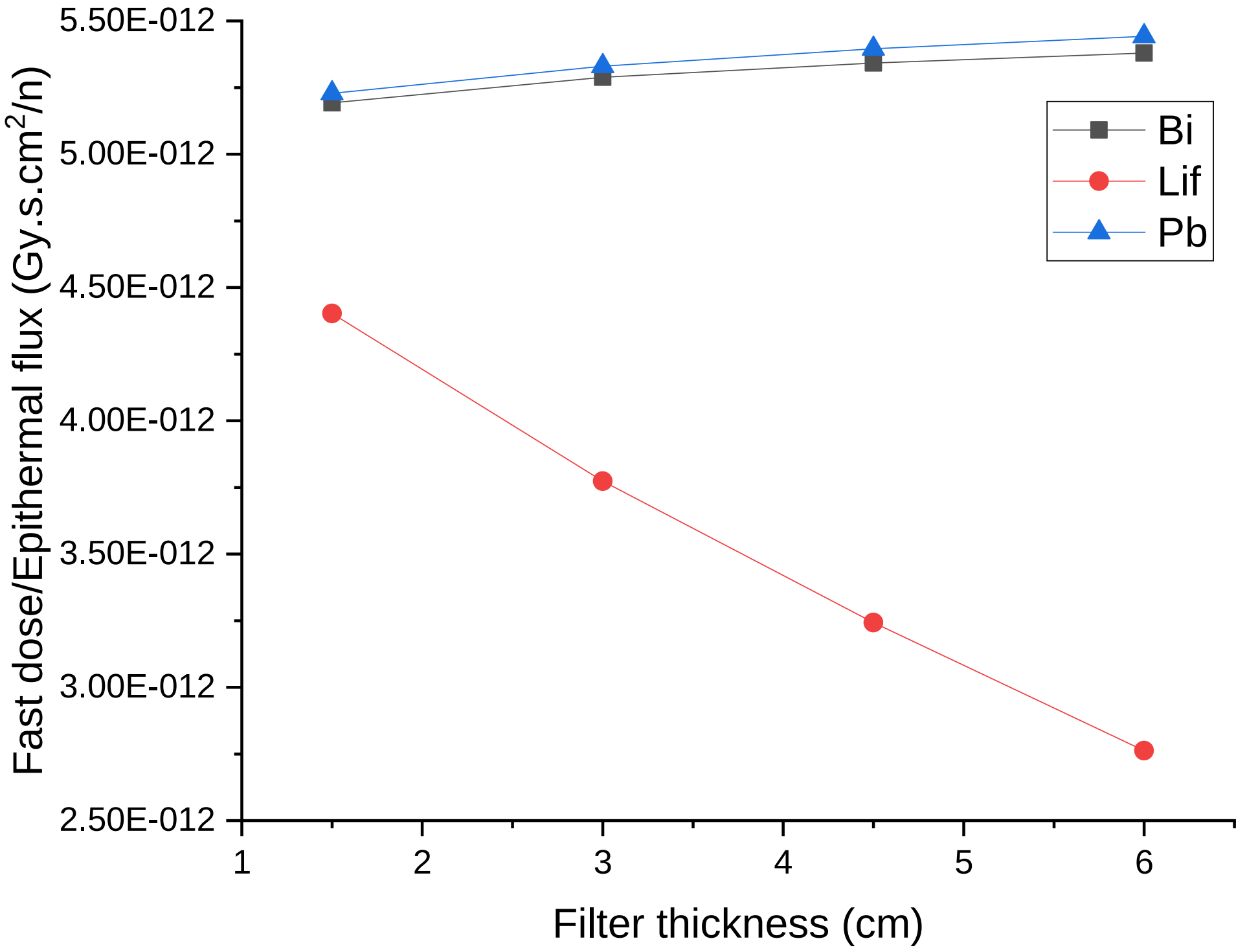

*Figure 7 Effect of different filter materials on fast dose to epithermal flux ratio of the BSA with moderator of 32 cm MgF2*

By using combinations of different filters the output can be improved as shown in Table 4.

| Pb (cm) | LiF (cm) | Bi (cm) | $\varphi_{epth}$ $n_{epith}.cm^{-2}s^{-1}$ | $\varphi_{ther}/\varphi_{epth}$ | $D_{fast}/\varphi_{epth}$ (Gy.cm$^2$) | $D_{\gamma}/\varphi_{epth}$ (Gy.cm$^2$) |
|---|---|---|---|---|---|---|
| 0.52 | 0 | 0.5 | 6.01E+05 | 1.79E-02 | 4.17E-12 | 6.71E-13 |
| 1.52 | 0 | 0.52 | 5.46E+05 | 1.86E-02 | 4.19E-12 | 4.46E-13 |
| 3 | 0 | 1.52 | 4.36E+05 | 1.88E-02 | 4.19E-12 | 1.71E-13 |
| 3 | 1.52 | 0 | 4.72E+05 | 2.41E-02 | 3.87E-12 | 4.82E-13 |

*Table 4 the effect of different combinations of filters on different parameter values (32cm MgF2 moderator).*

In this work, hundreds of different trials were made for different components of BSA. The best result from these trials is obtained when the moderator is 32 cm of $MgF_2$, the reflector is 76 cm of Pb, the filters are 3 cm Pb and 1.52 cm Bi, and the collimator is a cone frustum of Pb with height 10 cm and its aperture diameter is 12 cm as shown in row 4 of Table 4 . Current to epithermal flux ratio of this BSA is 0.87, which is a good result. With further trials and different combinations, the output may be improved.

Figure 8 provides a general evaluation of the clinical parameters. For this BSA output, it has been demonstrated that AD is 5.5 cm, AR is 2.29, and ADDR is 4.1 × $10^{-4}$ Gy.Eq/min. The ADDR decreases excessively because the neutron generator's initial output is insufficient. In order to obtain the recommended epithermal flux, the generator's initial neutron fluence needs to be 1.14 × $10^{13}$ n/sec. Achieving 1.14 × $10^{13}$ n/sec for neutron fluence will result in an ADDR of 0.47 Gy.Eq/min, meaning that the therapy can be completed in 30 minutes or less.

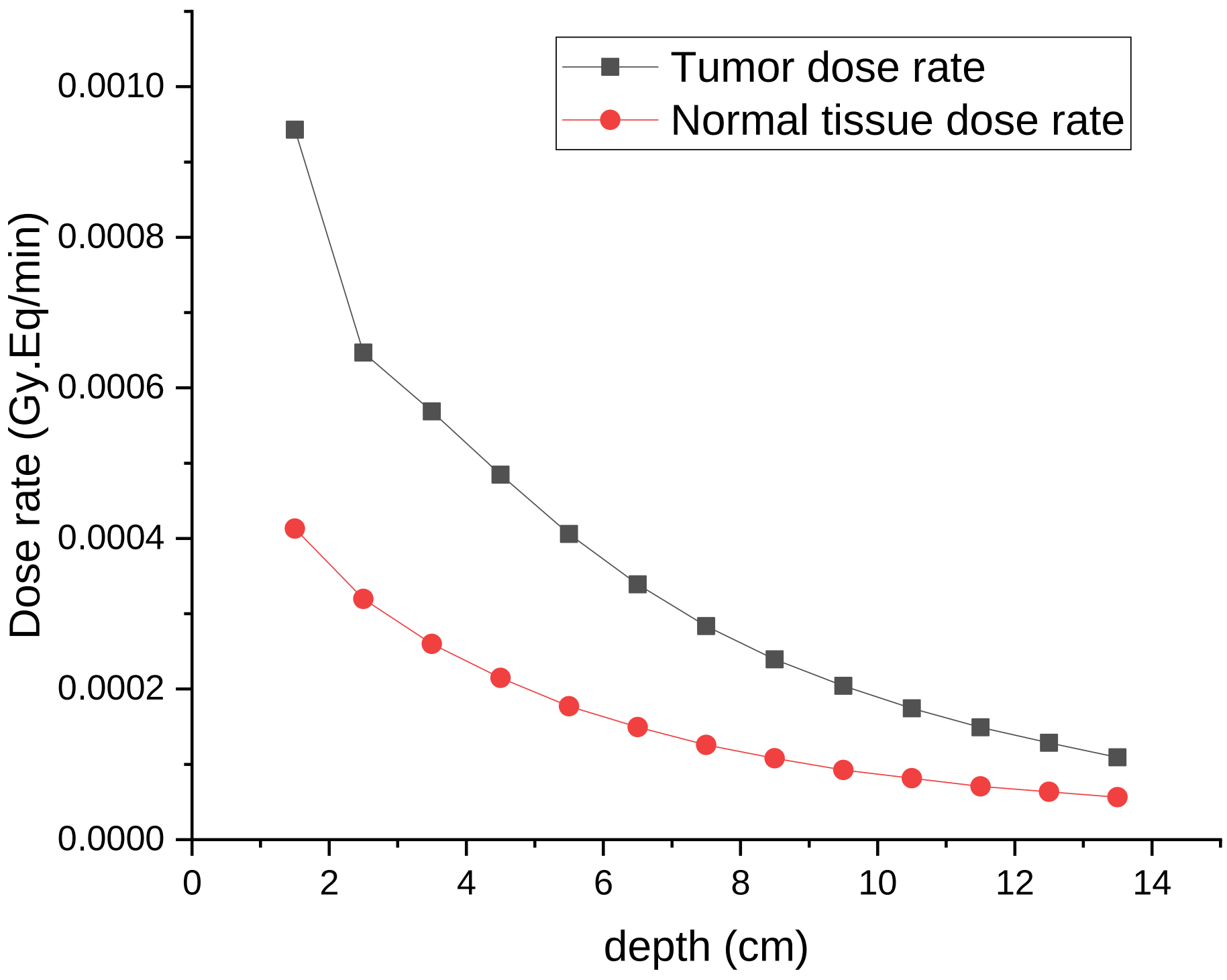

*Figure 8 Dose rate of tumor and healthy tissue in parallelepiped phantom*

## CONCLUSIONS

From the perspective of a medical physicist, BNCT entails a few steps: producing the necessary quantity of neutrons, adjusting them to the proper energy range, and figuring out the dosage deposition. This research leads to the conclusion that the DD-110 neutron generator emits less neutrons. Given that neutron generators are inexpensive, lightweight, and easy to construct and operate, more advancements in this field are encouraged. Further fast neutron dose filters should have more thickness in order to reduce the fast neutron dose. The epithermal flux drops with the addition of the fast neutron filter. Therefore, the right fast neutron dosage and epithermal flux can be obtained with a neutron generator with an output of 2E13 n/sec. Tables 5 and 6 compare the DD-109, the preceding neutron generator, with the DD serious.

| Generator model | $\varphi_{epth}$ ($n_{epith}.cm^{-2}s^{-1}$) | $\varphi_{ther}/\varphi_{epth}$ | $D_{fast}/\varphi_{epth}$ ($Gy.cm^2$) | $D_{\gamma}/\varphi_{epth}$ ($Gy.cm^2$) |
|---|---|---|---|---|
| DD-109[14] | 1E05 | 0.05 | 5.5E-13 | 2.49E-13 |
| DD-110 (proposed generator) | 4.36E+05 | 0.02 | 4.19E-12 | 1.71E-13 |

*Table 5 in-air parameter comparison of DD-109 and DD-110 neutron generators*

| Generator model | AD (cm) | AR | ADDR (Gy.Eq/min) | TT (min) |
|---|---|---|---|---|
| DD-109[14] | 12.1 | 3.7 | 0.31 | 40 |
| DD-110 (proposed generator) | 5.5 | 2.29 | 0.47 | 30 |

*Table 6 in-phantom parameters comparison of DD-109 and DD-110 neuron generators*

Broadly speaking, BNCT is a promising radiation treatment. One of the inputs in this treatment is a selective neutron source. Compared to research reactors and accelerators, neutron generators are safer, more affordable, and have a more straightforward design. On the other hand, neutron generators produce fewer neutrons than is advised. Therefore, the neutron generator needs to be improved even further.

Recently BNCT is in practice, the next challenge with this treatment modality is to make it simple and easy. The major aspect which makes this treatment modality simple and uncomplicated is good neutron source. Thus, one of the current research topics is obtaining neutron sources, such as DD neutron generators, that can create the right amount of neutrons.